\documentclass[a4paper, 12pt]{iopart}
\usepackage{latexsym}
\usepackage{iopams}
\usepackage{bbold}
\usepackage{tabls}
\usepackage{marvosym}
\usepackage{url}
\usepackage{graphicx}
\usepackage{epic,eepic}
\DeclareMathAlphabet{\tens}{OT1}{phv}{b}{n}
\begin{document}
\title{{\Huge Conservative discretization of the {\sc Einstein-Dirac} equations in spherically symmetric spacetime}}
\author{{\sc Benedikt Zeller} and {\sc Ralf Hiptmair}}
\date{20.03.2006}
\address{Department of Mathematics, ETH Zurich, CH-8092 Zurich, Switzerland}
\eads{\mailto{bzeller@math.ethz.ch}, \mailto{hiptmair@sam.math.ethz.ch}}
\begin{abstract}
In computational relativity, critical behaviour near the black hole threshold has been studied numerically for several models in the last decade. In this paper we present a spatial Galerkin method, suitable for finding numerical solutions of the {\sc Einstein-Dirac} equations in spherically symmetric spacetime (in polar/areal coordinates). The method features exact conservation of the total electric charge and allows for a spatial mesh adaption based on physical arclength. Numerical experiments confirm excellent robustness and convergence properties of our approach. Hence, this new algorithm is well suited for studying critical behaviour.
\end{abstract}
\maketitle
\pagestyle{plain}
\tableofcontents
\section{Introduction}
\subsection{Preface}
\label{sec:preface}
\paragraph{} In 1998-1999 {\sc F. Finster}, {\sc J. Smoller} and {\sc S.T. Yau} published static soliton-like solutions of the massive {\sc Einstein-Dirac} equations in a spherically symmetric spacetime in \cite{Finster_Felix_PLSolotEDE} and {\sc D.W. Schaefer}, {\sc  D.A. Steck} and {\sc J.F. Ventrella} studied the critical collapse of that system numerically. Although the applied iterative {\sc Crank-Nicholson} finite differencing scheme does neither conserve the total electric charge exactly nor make use of adaptive mesh grid refinement methods, the authors were able to investigate critical collapse for low ($m=0.25$) and moderate ($m=0.5$) particle masses for suitable initial data (See \cite{Schaefer_Steck_Ventrella_CritCollotEDF} for details).%
\paragraph{} Yet, these non adaptive finite difference methods fail for simulations of critical collapse for large particle masses ($m\geq 1$) or propagating {\sc Dirac} waves as initial data. Therefore we performed a {\sc Galerkin} discretization of the same equations and combined this with a physically motivated adaptive mesh grid refinement algorithm. In this paper we give a proof of the discretized conservation law, show some convergence tests and compare our results to those in \cite{Schaefer_Steck_Ventrella_CritCollotEDF}.
\paragraph{} Finally we note, that critical behaviour of the massless {\sc Einstein-Dirac} system has already been studied using iterative {\sc Crank-Nicholson} finite differencing with adaptive mesh grid refinement by {\sc J.F. Ventrella} and {\sc M.W. Choptuik} in \cite{Ventrella_Choptuik_CritPitEMDS}. 
%
\subsection{The basic model}
\label{sec:basic-model}
\paragraph{} Throughout this paper we use {\sc Planckian} units, where the speed of light, {\sc Planck}'s action and {\sc Newton}'s gravitational constant satisfy $c=\hbar=G_{\mathrm{N}}=1$. We consider a spherically symmetric spacetime in polar/areal coordinates $(t,r,\theta,\phi)$ with metric
\begin{eqnarray}\label{eq:AnsatzMetric}
\hspace{-1cm}{\tens g}=A(t,r)\,{\tens d}t\otimes{\tens d}t-B(t,r)\,{\tens
d}r\otimes{\tens d}r-r^{2}\,{\tens d}\theta\otimes{\tens
d}\theta-r^{2}\sin^{2}(\theta)\,{\tens d}\phi\otimes{\tens d}\phi,
\end{eqnarray}
where $A(t,r)$ and $B(t,r)$ are positive valued functions. We couple this metric by {\sc Einstein}'s equations to a massive $2$-fermion system of two spin-$\frac{1}{2}$ particles in singlet state. In order to find a coordinate representation of the {\sc Einstein-Dirac} equations, we constructed a complex $4\times 4$ matrix representation of the {\sc Clifford} bundle and an appropriate spin bundle with fibers isomorphic to $\mathbb{C}^{4}$, by using tetrad-formalism\footnote{We used {\sc Pauli-Dirac} representation throughout this paper, but one ends up with exactly the same expressions for all relevant physical quantities by using any other representation within the same unitary class, e.g. {\sc Weyl}'s or {\sc Majorana}'s representations \cite{Mandl_Shaw_QFT}.} \cite{Birrell_Davies_QFTiCS}, \cite{Finster_Felix_locU22SymiRQM}. This yields $\gamma$-matrices with $\{\gamma^{\mu},\gamma^{\nu}\}=2\,g^{\mu\nu}\,\mathbb{1}$, a representation of the $(2,2)$-signed semi-hermitian scalar product $\langle.,.\rangle$ on spinors and the connection coefficients of the {\sc Levi-Civita} connection on the spin bundle. In analogy to the ansatz for the {\sc Dirac} spinors in flat spacetime interacting with a spherically symmetric potential, we write the wavefunctions in the form\footnote{The prefactors $\frac{1}{2\,\sqrt{\pi}\,r\,B^{\frac{1}{4}}}$ are slightly different from those used in \cite{Finster_Felix_PLSolotEDE} to remove any time derivatives of $A$ and $B$ from {\sc Dirac}'s equation and to make the normalization condition for the functions $\alpha$ and $\beta$ as simple as possible.}
\numparts\begin{eqnarray}
\label{eq:AnsatzSpinors1}\fl
\psi_{1}&=&\frac{1}{2\,\sqrt{\pi}\,r\,B^{\frac{1}{4}}}\left[\begin{array}{c}\alpha\left[\begin{array}{c}1\\0\end{array}\right]\\[5pt]i\beta\sigma^{r}\left[\begin{array}{c}1\\0\end{array}\right]\end{array}\right]=\frac{1}{2\,\sqrt{\pi}\,r\,B^{\frac{1}{4}}}\left[\begin{array}{c}\alpha\\0\\
i\sin(\theta)\sin(\phi)\beta\\\left[\,i\cos(\theta)-\sin(\theta)\cos(\phi)\,\right]\beta\end{array}\right]\\[5pt]
\label{eq:AnsatzSpinors2}\fl
\psi_{2}&=&\frac{1}{2\,\sqrt{\pi}\,r\,B^{\frac{1}{4}}}\left[\begin{array}{c}\alpha\left[\begin{array}{c}0\\1\end{array}\right]\\[5pt]i\beta\sigma^{r}\left[\begin{array}{c}0\\1\end{array}\right]\end{array}\right]=\frac{1}{2\,\sqrt{\pi}\,r\,B^{\frac{1}{4}}}\left[\begin{array}{c}0\\\alpha\\\left[\,i\cos(\theta)+\sin(\theta)\cos(\phi)\,\right]\beta\\
-\,i\sin(\theta)\sin(\phi)\beta\end{array}\right],
\end{eqnarray}\endnumparts
where $\alpha(t,r)$ and $\beta(t,r)$ are complex valued functions. These wavefunctions satisfy
\numparts
\begin{eqnarray}
\psi_{i}^{\dagger}\psi_{j}&=&\frac{\delta_{ij}}{4\pi r^{2}\sqrt{B}}\left[\,|\alpha|^{2}+|\beta|^{2}\,\right]\\[5pt]\left\langle\psi_{i},\psi_{j}\right\rangle&=&\frac{\delta_{ij}}{4\pi r^{2}\sqrt{B}}\left[\,|\alpha|^{2}-|\beta|^{2}\,\right].
\end{eqnarray}
\endnumparts 
\paragraph{} On each spacelike slice of constant $t$, denoted by $\Sigma(t)$, we construct the fermion state of the two particles, given by
\begin{eqnarray}\label{eq:FermionState}
\left|\Psi\right\rangle=\frac{1}{\sqrt{2}}\left[\,\left|\psi_{1}\right\rangle\otimes\left|\psi_{2}\right\rangle-\left|\psi_{2}\right\rangle\otimes\left|\psi_{1}\right\rangle\,\right].
\end{eqnarray}
Assuming each particle carries\footnote{We don't couple the particles to the electromagnetic field in this paper. The particle charge $q$ is introduced just for convenience.} an electric charge $q$, the components of the charge current density vector ${\tens J}=q\sum_{p=1}^{2}\left\langle\psi_{p},\gamma^{\mu}\psi_{p}\right\rangle\,\tens{e}_{\mu}=J^{\mu}\,\tens{e}_{\mu}$ would be
\begin{eqnarray}\label{eq:ChargeCurrDensVec}
\left[J^{\mu}\right]=\frac{q}{2\pi r^{2}}
\left[\begin{array}{c}\frac{1}{\sqrt{AB}}\left[\,|\alpha|^2+|\beta|^{2}\,\right]\\\frac{2}{B}\,\mathrm{Im}\left(\alpha\beta^{*}\right)\\
0\\0\end{array}\right]
\end{eqnarray}
and the total electric charge on each slice $\Sigma(t)$ sums up to
\begin{eqnarray}\label{eq:TotElectricCharge}
\hspace{-1.5cm}Q:=\int_{\Sigma}\left\langle\,{\tens J},{\tens N}\,\right\rangle d\Sigma=q\int_{\Sigma}\left[\,\psi_{1}^{\dagger}\psi_{1}+\psi_{2}^{\dagger}\psi_{2}\,\right]d\Sigma=2q\int_{0}^{\infty}\left[\,|\alpha|^2+|\beta|^{2}\,\right]dr\stackrel{!}{=}2q,
\end{eqnarray}
which implies the normalization \cite{Finster_Felix_PLSolotEDE} condition\footnote{The same result can also be optained by normalizing the quantum states on each slice $\Sigma(t)$, i.e. by requiring $\left\langle\psi_{p}\vert\psi_{p}\right\rangle=\left\langle\Psi\vert\Psi\right\rangle=1$.}
\begin{eqnarray}\label{eq:NormCond}
\int_{0}^{\infty}\left[\,|\alpha|^2+|\beta|^{2}\,\right]dr=1.
\end{eqnarray}
With the ansatz (\ref{eq:AnsatzSpinors1},\ref{eq:AnsatzSpinors2}) the equations of motion for the wavefunctions $\psi_{1}$ and $\psi_{2}$ are independent of each other, i.e. they satisfy {\sc Dirac}'s equation separately:
\begin{eqnarray}\label{eq:DiracGeom}
\left[\,i\mathcal{D}-m\,\mathbb{1}\,\right]\psi_{p}=0
\end{eqnarray}
Here $m>0$ is the particle mass and
$\mathcal{D}:=\gamma^{\sigma}\boldsymbol{\nabla}_{{\tens e}_{\sigma}}$ denotes the {\sc Dirac} operator constructed from the metric (\ref{eq:AnsatzMetric}). The stress energy tensor of the system is just the sum of the stress energies of both wavefunctions, its components are given by
\begin{eqnarray}\label{eq:StressEnergyGeom}
T_{\mu\nu}=\frac{i}{2}\sum_{p=1}^{2}\left[\,\left\langle\,\psi_{p},\gamma_{(\mu}\boldsymbol{\nabla}_{{\tens e}_{\nu)}}\psi_{p}\right\rangle-\left\langle\,\boldsymbol{\nabla}_{{\tens e}_{(\mu}}\psi_{p},\gamma_{\nu)}\psi_{p}\right\rangle\,\right].
\end{eqnarray}
From the ansatz (\ref{eq:AnsatzSpinors1},\ref{eq:AnsatzSpinors2}) we get
\begin{eqnarray}\label{eq:StressEnergyCoord}
\hspace{-1.5cm}T_{00}&=&\frac{1}{2\pi
r^{2}}\,\sqrt{\frac{A}{B}}\,\mathrm{Im}\left(\alpha{\dot\alpha}^{*}+\beta{\dot\beta}^{*}\right)\\[5pt]
\hspace{-1.5cm}T_{01}&=&\frac{1}{4\pi r^{2}}\,\mathrm{Re}\left(\alpha^{*}{\dot\beta}-{\dot\alpha}\beta^{*}\right)+\frac{1}{4\pi r^{2}}\,\sqrt{\frac{A}{B}}\,\mathrm{Im}\left(\alpha{\alpha'}^{*}+\beta{\beta'}^{*}\right)=T_{10}\nonumber\\[5pt]
\hspace{-1.5cm}T_{11}&=&\frac{1}{2\pi r^{2}}\,\mathrm{Re}\left(\alpha^{*}\beta'-\alpha'\beta^{*}\right)\nonumber\\[5pt]
\hspace{-1.5cm}T_{22}&=&\frac{1}{2\pi r\,\sqrt{B}}\,\mathrm{Re}\left(\alpha^{*}\beta\right)\ \mathrm{and}\ T_{33}=\frac{1}{2\pi r\,\sqrt{B}}\,\mathrm{Re}\left(\alpha^{*}\beta\right)\sin^{2}(\theta).\nonumber
\end{eqnarray}
\paragraph{} Plugging the ansatz (\ref{eq:AnsatzSpinors1},\ref{eq:AnsatzSpinors2}) into (\ref{eq:DiracGeom}) yields two first order evolution equations, which are linear in $\alpha$ and $\beta$:
\numparts\begin{eqnarray}
\label{eq:alphat}{\dot\alpha}&=&-\,i\,\sqrt{\frac{A}{B}}\,\beta'-\frac{i}{2}\left[\sqrt{\frac{A}{B}}\,\right]'\beta-\frac{i}{r}\,\sqrt{A}\,\beta-im\,\sqrt{A}\,\alpha\\[5pt]
\label{eq:betat}{\dot\beta}&=&+\,i\,\sqrt{\frac{A}{B}}\,\alpha'+\frac{i}{2}\left[\sqrt{\frac{A}{B}}\,\right]'\alpha-\frac{i}{r}\,\sqrt{A}\,\alpha+im\,\sqrt{A}\,\beta
\end{eqnarray}\endnumparts
{\sc Einstein}'s equations involving the stress energy (\ref{eq:StressEnergyGeom}) leads to the two ordinary differential equations in the radial variable
\numparts\begin{eqnarray}
\label{eq:Ar}\frac{A'}{A}&=&\frac{1}{r}\left[\,B-1+2\left[\,\alpha^{*}\beta'-\alpha'\beta^{*}+\alpha{\beta^{*}}'-{\alpha^{*}}'\beta\,\right]\,\right]\\[5pt]
\label{eq:Br}\frac{B'}{B}&=&\frac{1}{r}\left[\,1-B+2\left[\,\alpha^{*}\beta'-\alpha'\beta^{*}+\alpha{\beta^{*}}'-{\alpha^{*}}'\beta\,\right]\,\right]\\[5pt]
{}&{}&+\,\frac{4m}{r}\,\sqrt{B}\,\left[\,|\alpha|^2-|\beta|^{2}\,\right]+\frac{4}{r^{2}}\,\sqrt{B}\,\left[\,\alpha\beta^{*}+\alpha^{*}\beta\,\right],\nonumber
\end{eqnarray}\endnumparts
which could be identified as the {\sc Hamiltonian} constraint and the slicing condition in spherical symmetry, and an evolution equation for $B$:
\begin{eqnarray}
\label{eq:Bt}\frac{\dot B}{B}=\frac{2i}{r}\,\sqrt{\frac{A}{B}}\left[\,\alpha'\alpha^{*}-\alpha{\alpha^{*}}'+\beta'\beta^{*}-\beta{\beta^{*}}'\,\right]
\end{eqnarray}
The equations (\ref{eq:Br}) and (\ref{eq:Bt}) are consistent. One easily checks that the evolution of $B$ w.r.t. (\ref{eq:Bt}) yields the same as solving (\ref{eq:Br}) on each slice $\Sigma(t)$. So one of them is redundant.
%
\subsection{The evolution boundary value problem}
\label{sec:init-bound-value}
\paragraph{} We are looking for regular and asymptotically flat solutions of (\ref{eq:alphat},\ref{eq:betat}) and (\ref{eq:Ar},\ref{eq:Br},\ref{eq:Bt}). Let $T>0$, $I:=[\,0,T\,]$ and $J:=[\,0,\infty)$. We seek solutions of (\ref{eq:alphat},\ref{eq:betat}) and (\ref{eq:Ar},\ref{eq:Br},\ref{eq:Bt}), such that for all $t\in I$ the variables $\alpha$ and $\beta$ as functions of $r$ are in the {\sc Sobolev} space $W^{1,\infty}(J)\cap H^{1}(J)$ and $A$ and $B$ as functions of $r$ are elements of $W^{1,\infty}(J)$. This leads us to the ``boundary conditions''
\begin{eqnarray}
\lim_{r\to \infty}A(t,r)=\lim_{r\to \infty}B(t,r)=\lim_{r\to 0}B(t,r)=1\ \forall\,t\in I.
\end{eqnarray}
Because of the factor $\frac{1}{r}$ in the ansatz (\ref{eq:AnsatzSpinors1},\ref{eq:AnsatzSpinors2}), $\alpha$ and $\beta$ have to fulfill
\begin{eqnarray}
\alpha(t,r)=O(r)\ \mathrm{and}\ \beta(t,r)=O(r)\ \forall\,t\in I\ \mathrm{for}\ r\to 0.
\end{eqnarray}
For the numerical treatment of our problem, and in order to guarantee the asymptotical flatness of the solution, we consider the {\sc Dirac} matter to be concentrated inside a ball of coordinate radius $R>0$ around the center of spherical symmetry. Due to {\sc Birkhoff}'s theorem, the part of spacetime at regions where $r>R$ has to be {\sc Schwarzschild} \cite{Wald_Robert_GR}. Therefore we consider the following boundary/initial value problem on $I\times J_{R}$ with $J_{R}:=[\,0,R\,]$ for $\alpha$, $\beta$, $A$ and $B$:
\begin{eqnarray}\label{eq:InitValProb}
\fl \left\{\,\begin{array}{rcl|l} \displaystyle \mathrm{PDEs:\ (\ref{eq:alphat},\ref{eq:betat})}&{}&\ &\ \mathrm{for}\ (t,r)\in(0,T)\times(0,R)\\[5pt] \displaystyle \mathrm{ODEs:\ (\ref{eq:Ar},\ref{eq:Br})}&{}&\ &\ \mathrm{for}\ (t,r)\in I\times (0,R)\\[5pt]
\displaystyle\alpha(t,0)=\beta(t,0)=\alpha(t,R)=\beta(t,R)&=&0\ &\ \mathrm{for}\ t\in I\\[5pt]
\displaystyle B(t,0)&=&1\ &\ \mathrm{for}\ t\in I\\[5pt]
\displaystyle A(t,R)&=&\frac{1}{B(t,R)}\ &\ \mathrm{for}\ t\in I\\[5pt]
\displaystyle \alpha(0,r)&=&\alpha_{0}(r)\ &\ \mathrm{for}\ r\in J_{R}\\[5pt]
\displaystyle\beta(0,r)&=&\beta_{0}(r)\ &\ \mathrm{for}\ r\in J_{R}
\end{array}\right.
\end{eqnarray}
Here $\alpha_{0},\beta_{0}\in W^{1,\infty}(J_{R})\cap H^{1}_{0}(J_{R})$ denotes some suitable initial data. Because of the boundary conditions for $\alpha$ and $\beta$, outgoing {\sc Dirac} waves are reflected at $r=R$, which is an unphysical requirement. For initial data with a fast decay (i.e. exponential decay) for large $r$ and by specifying $R$ large enough, these reflected waves do not interfere with what is going on in the region of interest, that is, for small $r$.
%
\subsection{Conservation laws}
\label{sec:conservation-laws}
\paragraph{} The total electric charge from (\ref{eq:TotElectricCharge}), i.e. the normalization condition (\ref{eq:NormCond}), and the total ADM-mass of spacetime are constants of motion. For the charge this can be seen by a straightforward computation \cite{Finster_Felix_PersComm} using (\ref{eq:alphat},\ref{eq:betat}):
\begin{eqnarray}
\hspace{-1.5cm}{\dot
Q}&=&\frac{d}{dt}\int_{0}^{\infty}\left[\,\left\vert\alpha\right\vert^{2}+
\left\vert\beta\right\vert^{2}\,\right]dr=2\,\mathrm{Re}
\int_{0}^{\infty}\left[\,{\dot\alpha}\alpha^{*}+{\dot\beta}\beta^{*}\,\right]dr\\[7pt]
\hspace{-1.5cm}{}&=&2\,\mathrm{Re}\int_{0}^{\infty}\left[-\,i\,\sqrt{\frac{A}{B}}\,\beta'\alpha^{*}-\,\frac{i}{2}\left[\,\sqrt{\frac{A}{B}}\,\right]'\,\beta\alpha^{*}+\,i\,\sqrt{\frac{A}{B}}\,\alpha'\beta^{*}+\,\frac{i}{2}\left[\,\sqrt{\frac{A}{B}}\,\right]'\,\alpha\beta^{*}\right.\nonumber\\[7pt]
\hspace{-1.5cm}{}&\!\!\!{}\!\!\!&\displaystyle\left.-\,i\,\frac{\sqrt{A}}{r}\left[\,\alpha\beta^{*}+\alpha^{*}\beta\,\right]+im\sqrt{A}\,\left[\,\left\vert\beta\right\vert^{2}-\left\vert\alpha\right\vert^{2}\,\right]\right]dr\nonumber\\[7pt]
\hspace{-1.5cm}{}&=&2\,\mathrm{Re}\,i\int_{0}^{\infty}\left[\,\sqrt{\frac{A}{B}}\left[\,\alpha'\beta^{*}-\beta'\alpha^{*}\,\right]+\frac{1}{2}\left[\,\sqrt{\frac{A}{B}}\,\right]'\left[\,\alpha\beta^{*}-\beta\alpha^{*}\,\right]\right]dr\nonumber\\[7pt]
\hspace{-1.5cm}{}&=&2\,\mathrm{Re}\,i\int_{0}^{\infty}\left[\,\sqrt{\frac{A}{B}}\left[\,\alpha'\beta^{*}-\beta'\alpha^{*}\,\right]-\frac{1}{2}\,\sqrt{\frac{A}{B}}\left[\,\alpha'\beta^{*}+\alpha{\beta^{*}}'-\beta'\alpha^{*}-\beta{\alpha^{*}}'\,\right]\right]dr\nonumber\\[7pt]
\hspace{-1.5cm}{}&=&\mathrm{Re}\,i\int_{0}^{\infty}\sqrt{\frac{A}{B}}\left[\,\alpha'\beta^{*}-\beta'\alpha^{*}-\alpha{\beta^{*}}'+\beta{\alpha^{*}}'\,\right]dr\nonumber\\[7pt]
\hspace{-1.5cm}{}&=&2\,\mathrm{Re}\,i\int_{0}^{\infty}\sqrt{\frac{A}{B}}\,\mathrm{Re}\left(\,\alpha'\beta^{*}-\beta'\alpha^{*}\right)dr=0\nonumber
\end{eqnarray}
The conservation of the ADM-mass of spacetime is obvious from the following formula, which could be transformed into a conserved integral by using {\sc Einstein}'s equations and (\ref{eq:StressEnergyCoord}):
\begin{eqnarray}
M_{\mathrm{ADM}}(\Sigma)=\frac{1}{2}\,\lim_{r\to\infty}r\left[\,1-\frac{1}{B}\,\right]=2\int_{0}^{\infty}\mathrm{Im}\left(\alpha{\dot\alpha}^{*}+\beta{\dot\beta}^{*}\right)\frac{1}{\sqrt{AB}}\,dr
\end{eqnarray}
%
\section{Conservative discretization}
\label{sec:cons-disc}
\paragraph{} To find a conservative {\sc Galerkin} discretization, we first consider the model problem
\begin{eqnarray}\label{eq:ModelProblem}
{\dot\alpha}&=&-\,if\beta'-\frac{i}{2}\,f'\beta-ig\alpha-\frac{i}{r}\,h\beta\\[5pt]
{\dot\beta}&=&+\,if\alpha'+\frac{i}{2}\,f'\alpha+ig\beta-\frac{i}{r}\,h\alpha\;,\nonumber
\end{eqnarray}
where $f,g,h\in C^{1}\left(I,H^{1}(J_{R})\right)$ are real valued. We seek complex valued solutions $\alpha(t),\beta(t)\in H^{1}_{0}(J_{R})$. In the sequel $(.,.)$ denotes the scalar product in $L^{2}(J_{R})$. The weak formulation of (\ref{eq:ModelProblem}) reads
\begin{eqnarray}\label{eq:ModelProblemWeak}
\left({\dot\alpha},v\right)&=&\left(-\,if\beta'-\frac{i}{2}\,f'\beta-ig\alpha-\frac{i}{r}\,h\beta,v\right)\\[5pt]
\left({\dot\beta},w\right)&=&\left(+\,if\alpha'+\frac{i}{2}\,f'\alpha+ig\beta-\frac{i}{r}\,h\alpha,w\right),\nonumber
\end{eqnarray}
for all $v,w\in H^{1}_{0}(J_{R})$. In the spirit of the method of lines, we first carry out the spatial {\sc Galerkin} semi-discretization of (\ref{eq:ModelProblemWeak}). It can be obtained by replacing the function space $H^{1}_{0}(J_{R})$ with a finite dimensional subspace $V_{n}$. Please note that these subspaces have to contain continuous functions that vanish for $r=0$ and $r=R$. As a basis for $V_{n}$ we choose real valued global shape functions $e_{i}\in H_{0}^{1}(J_{R})$ for $i\in\{1,\ldots, n\}$, $n:=\mathrm{dim}\,V_{n}\in\mathbb{N}$. Thus we can represent spatially discrete approximations of $\alpha$ and $\beta$ in (\ref{eq:ModelProblemWeak}) by
\begin{eqnarray}\label{eq:Repralphabeta}
\alpha(t,r)\approx\sum_{i=1}^{n}\alpha^{i}(t)\,e_{i}(r)\quad \mathrm{and}\quad \beta(t,r)\approx
\sum_{i=1}^{n}\beta^{i}(t)\,e_{i}(r)\;.
\end{eqnarray}
Next, we define the $n\times n$ {\sc Galerkin}-matrices $E$, $F$, $G$ and $H$ with components
\begin{eqnarray}
\fl E_{ij}:=\left(e_{i},e_{j}\right),\
F_{ij}:=\left(fe'_{i}+\frac{1}{2}\,f'e_{i},e_{j}\right),\
G_{ij}:=\left(ge_{i},e_{j}\right)\ \mathrm{and}\
H_{ij}:=\left(\frac{1}{r}\,he_{i},e_{j}\right).
\end{eqnarray}
From the integration by parts formula we get
\begin{eqnarray}
\fl
F_{ij}&=&\int_{0}^{R}\left[\,fe'_{i}e_{j}+\frac{1}{2}\,f'e_{i}e_{j}\right]dr=-\int_{0}^{R}\left[\,fe_{i}e'_{j}+\frac{1}{2}\,f'e_{i}e_{j}\right]dr=-\,F_{ji}
\end{eqnarray}
and the above matrices obey the algebraic properties
\begin{eqnarray}\label{eq:MatrAlgProp}
E^{\dagger}=E,\ F^{\dagger}=-\,F,\ G^{\dagger}=G\ \mathrm{ and }\ H^{\dagger}=H\;.
\end{eqnarray}
\paragraph{} Then, we tackle discretization in time: We recall that (\ref{eq:ModelProblem}) has the character of a wave equation with a quadratic first integral, the total electric charge $Q$, see Sect.~\ref{sec:conservation-laws}. Exact conservation of quadratic first integrals is guaranteed by the so-called implicit midpoint rule, the simplest representative of {\sc Gauss} collocation {\sc Runge-Kutta} timestepping methods \cite[Sect.~IV.2.1]{Hairer_Lubich_Wanner_GeomNumInt}. Formally, this timestepping scheme is second order accurate with respect to the size of the timestep. To describe its application to the semi-discrete system, we fix two instances in time $0\leq t_{m} < t_{m+1} \leq T$, introduce the (local) timestep $\tau:=t_{m+1}-t_{m}$, the notations
\begin{eqnarray}
\alpha_{m}:=\left[\begin{array}{c}\alpha^{1}(t_{m})\\\vdots\\\alpha^{n}(t_{m})\end{array}\right],\
\Delta\alpha:=\alpha_{m+1}-\alpha_{m},\
{\bar\alpha}:=\frac{1}{2}\left[\,\alpha_{m+1}+\alpha_{m}\,\right]
\end{eqnarray}
and analogous for $\beta_{m}$, $\Delta\beta$ and ${\bar\beta}$. The fully discrete version of (\ref{eq:ModelProblemWeak}) then reads
\begin{eqnarray}\label{eq:ModelProblemDisc}
\frac{1}{\tau}\,E\,\Delta\alpha&=&-\,iF{\bar\beta}-iG{\bar\alpha}-iH{\bar\beta}\\[5pt]
\frac{1}{\tau}\,E\,\Delta\beta&=&+\,iF{\bar\alpha}+iG{\bar\beta}-iH{\bar\alpha}.\nonumber
\end{eqnarray}
Obviously, each timestep involves the solution of a linear system of equations. The discrete total electric charge is given by
\begin{eqnarray}
Q_{m}=2q\left[\,\alpha^{\dagger}_{m}E\alpha_{m}+\beta^{\dagger}_{m}E\beta_{m}\,\right].
\end{eqnarray}
Because of the properties (\ref{eq:MatrAlgProp}), it is conserved exactly along the time evolution generated by the discrete equations (\ref{eq:ModelProblemDisc}):
\begin{eqnarray}
\hspace{-1.5cm}\frac{1}{2q\tau}\,\Delta Q
&=&{\bar\alpha}^{\dagger}A\Delta\alpha+{\bar\beta}^{\dagger}A\Delta\beta+i\underbrace{2\,\mathrm{Im}\left(\alpha_{m}^{\dagger}A\alpha_{m+1}+\beta_{m}^{\dagger}A\beta_{m+1}\right)}_{=:\,S\in\mathbb{R}}\\[5pt]
\hspace{-1.5cm}{}&=&i\underbrace{\left[\,{\bar\beta}^{\dagger}F{\bar\alpha}-{\bar\alpha}^{\dagger}F{\bar\beta}\,\right]}_{\in\mathbb{R}}+i\underbrace{\left[\,{\bar\alpha}^{\dagger}H{\bar\beta}+{\bar\beta}^{\dagger}H{\bar\alpha}\,\right]}_{\in\mathbb{R}}+i\underbrace{\left[\,{\bar\beta}^{\dagger}G{\bar\beta}-{\bar\alpha}^{\dagger}G{\bar\alpha}\,\right]}_{\in\mathbb{R}}+iS\nonumber\\[5pt]
\hspace{-1.5cm}{}&=&0.\nonumber
\end{eqnarray}
We point out, that this conservation property holds for any choice of trial/test space $V_{n}$, as long as the same $V_{n}$ is used for every timestep.
%
\section{Computer implementation}
\label{sec:comp-impl}
%
\subsection{Variables}
\label{sec:variables}
\paragraph{} We implemented the discretization scheme described above for the six real variables $X_{a}$, $Y_{a}$, $X_{b}$, $Y_{b}$, $a$ and $b$, defined by
\begin{eqnarray}
\alpha=X_{a}+i\,Y_{a},\ \beta=X_{b}+i\,Y_{b},\ A=\mathrm{e}^{a}\ \mathrm{and}\
B=\mathrm{e}^{b}\;,
\end{eqnarray}
where the representation of $A$ and $B$ is motivated by the essential positivity of the metric coefficient and the presence of logarithmic derivatives in (\ref{eq:Ar}), (\ref{eq:Br}) and (\ref{eq:Bt}). In these variables {\sc Dirac}'s equations read
\numparts\begin{eqnarray}
\label{eq:Xat}{\dot X}_{a}&=&+\,\mathrm{e}^{\frac{a-b}{2}}\,Y'_{b}+\frac{1}{2}\,\left[\mathrm{e}^{\frac{a-b}{2}}\right]'Y_{b}+\mathrm{e}^{\frac{a}{2}}\left[\,\frac{1}{r}\,Y_{b}+m\,Y_{a}\,\right]\\[5pt]
\label{eq:Yat}{\dot Y}_{a}&=&-\,\mathrm{e}^{\frac{a-b}{2}}\,X'_{b}+\frac{1}{2}\,\left[\mathrm{e}^{\frac{a-b}{2}}\right]'X_{b}-\mathrm{e}^{\frac{a}{2}}\left[\,\frac{1}{r}\,X_{b}+m\,X_{a}\,\right]\\[5pt]
\label{eq:Xbt}{\dot X}_{b}&=&-\,\mathrm{e}^{\frac{a-b}{2}}\,Y'_{a}+\frac{1}{2}\,\left[\mathrm{e}^{\frac{a-b}{2}}\right]'Y_{a}+\mathrm{e}^{\frac{a}{2}}\left[\,\frac{1}{r}\,Y_{a}-m\,Y_{b}\,\right]\\[5pt]
\label{eq:Ybt}{\dot Y}_{b}&=&+\,\mathrm{e}^{\frac{a-b}{2}}\,X'_{a}+\frac{1}{2}\,\left[\mathrm{e}^{\frac{a-b}{2}}\right]'X_{a}-\mathrm{e}^{\frac{a}{2}}\left[\,\frac{1}{r}\,X_{a}-m\,X_{b}\,\right]
\end{eqnarray}\endnumparts
and {\sc Einstein}'s equations are
\numparts\begin{eqnarray}
\label{eq:ar}a'&=&\frac{1}{r}\left[\,\mathrm{e}^{b}-1\,\right]+\frac{4}{r}\left[\,X_{a}X'_{b}-X'_{a}X_{b}+Y_{a}Y'_{b}-Y'_{a}Y_{b},\right]\\[5pt]
\label{eq:br}b'&=&\frac{1}{r}\left[\,1-\mathrm{e}^{b}\,\right]+\frac{4}{r}\left[\,X_{a}X'_{b}-X'_{a}X_{b}+Y_{a}Y'_{b}-Y'_{a}Y_{b},\right]\\[5pt]
{}&{}&+\,\frac{4m}{r}\,\mathrm{e}^{\frac{b}{2}}\left[\,X^{2}_{a}+Y^{2}_{a}-X^{2}_{b}-Y^{2}_{b}\,\right]+\frac{8}{r^{2}}\,\mathrm{e}^{\frac{b}{2}}\left[\,X_{a}X_{b}+Y_{a}Y_{b}\,\right]\nonumber
\end{eqnarray}\endnumparts
and
\begin{eqnarray}
\label{eq:bt}{\dot b}=\frac{4}{r}\,\mathrm{e}^{\frac{a-b}{2}}\left[\,X'_{a}Y_{a}-X_{a}Y'_{a}+X'_{b}Y_{b}-X_{b}Y'_{b}\,\right].
\end{eqnarray}
%
\subsection{Initial conditions}
\label{sec:initial-conditions}
\paragraph{} We implemented two classes of initial conditions. The one parameter family $(\Sigma)$ of {\sc Gaussian}'s, which was also used in \cite{Schaefer_Steck_Ventrella_CritCollotEDF}, is defined by
\begin{eqnarray}\label{eq:InitCondVen}
X_{a}(0,r)&:=&\left[\frac{2}{\pi}\right]^{\frac{1}{4}}\Sigma^{-\,\frac{3}{2}}\,r\,
\mathrm{e}^{-\,\frac{r^{2}}{4\Sigma^{2}}}\\[5pt]
Y_{a}(0,r)&:=&X_{b}(0,r)=Y_{b}(0,r)=0.\nonumber
\end{eqnarray}
Further we considered the $8$ parameter family $\left(\Sigma_{a},\Sigma_{b},P_{a},P_{b},\Psi_{a},\Psi_{b},\Psi_{w},p\right)$:
\begin{eqnarray}\label{eq:InitCondStd}
X_{a}(0,r)&:=&\frac{2^{\frac{1}{4}-\frac{p}{2}}\cos\left(\Psi_{w}\right)}{\sqrt{\Gamma\left(p+\frac{1}{2}\right)}\,\Sigma_{a}^{p+\frac{1}{2}}}\,\cos\left(\Psi_{a}+\frac{2\pi
r}{P_{a}}\right)\,r^{p}\,\mathrm{e}^{-\,\frac{r^{2}}{4\Sigma^{2}_{a}}}\\[5pt]
Y_{a}(0,r)&:=&\frac{2^{\frac{1}{4}-\frac{p}{2}}\cos\left(\Psi_{w}\right)}{\sqrt{\Gamma\left(p+\frac{1}{2}\right)}\,\Sigma_{a}^{p+\frac{1}{2}}}\,\sin\left(\Psi_{a}+\frac{2\pi
r}{P_{a}}\right)\,r^{p}\,\mathrm{e}^{-\,\frac{r^{2}}{4\Sigma^{2}_{a}}}\nonumber\\[5pt]
X_{b}(0,r)&:=&\frac{2^{\frac{1}{4}-\frac{p}{2}}\sin\left(\Psi_{w}\right)}{\sqrt{\Gamma\left(p+\frac{1}{2}\right)}\,\Sigma_{b}^{p+\frac{1}{2}}}\,\cos\left(\Psi_{b}+\frac{2\pi
r}{P_{b}}\right)\,r^{p}\,\mathrm{e}^{-\,\frac{r^{2}}{4\Sigma^{2}_{b}}}\nonumber\\[5pt]
Y_{b}(0,r)&:=&\frac{2^{\frac{1}{4}-\frac{p}{2}}\sin\left(\Psi_{w}\right)}{\sqrt{\Gamma\left(p+\frac{1}{2}\right)}\,\Sigma_{b}^{p+\frac{1}{2}}}\,\sin\left(\Psi_{b}+\frac{2\pi
r}{P_{b}}\right)\,r^{p}\,\mathrm{e}^{-\,\frac{r^{2}}{4\Sigma^{2}_{b}}}.\nonumber
\end{eqnarray}
Since critical behaviour is always studied by using a one parameter family (see e.g. \cite{Gundlach_Carsten_Cpigc}), one of the $8$ parameters will be chosen to vary, while the other $7$ remain fixed.
%
\subsection{Mesh and shape functions}
\label{sec:mesh-shape-functions}
\begin{figure}[htp]
\begin{center}
\begin{picture}(16,7)
\put(0,0){\includegraphics{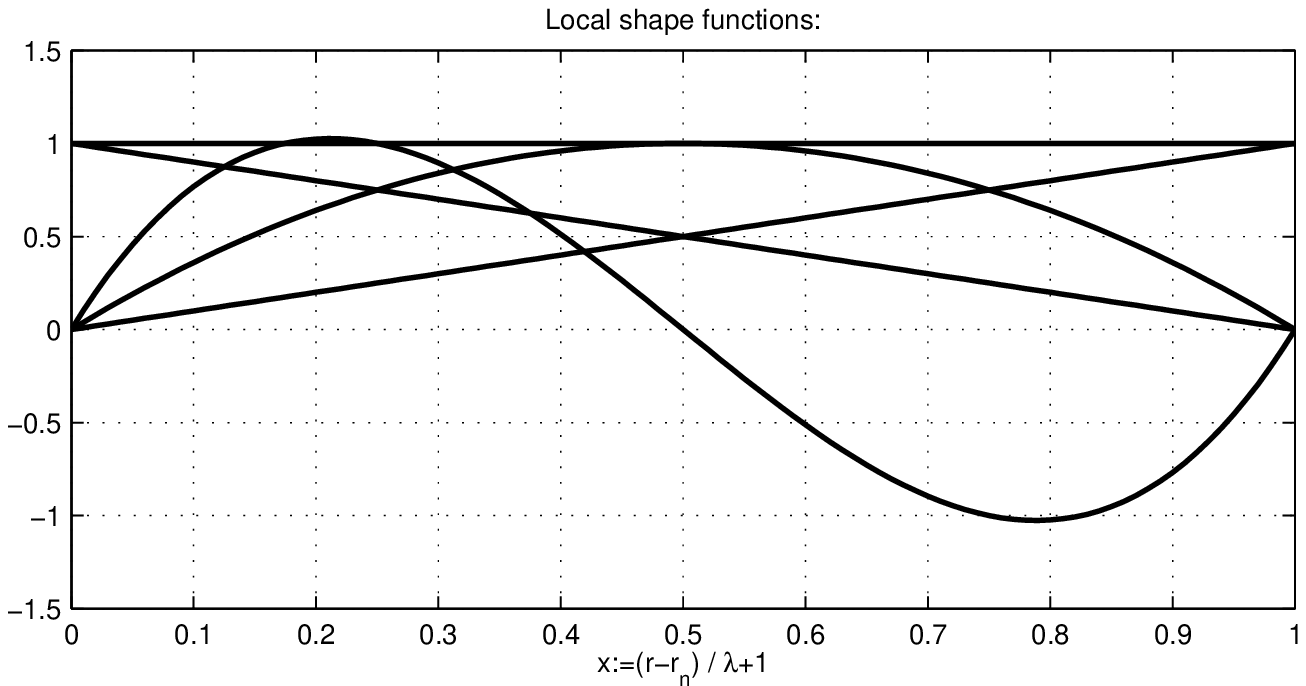}}
\end{picture}
{\bf \caption{\label{fig:plot-01}}\footnotesize {\textmd { Local shape functions of different polynomial degree on an interval of the spatial grid.}}}
\end{center}
\end{figure}

\paragraph{} As {\sc Galerkin} trial and test spaces for $X_{a}$, $Y_{a}$, $X_{b}$, and $Y_{b}$ we rely on spaces of piecewise polynomial continuous functions on a spatial mesh covering $J_{R}$. In other words, we use classical {\sc Lagrangian} finite element spaces. More precisely, the mesh is a not necessarily equidistant grid on $J_{R}$, whose $N$ cells, $N\in\mathbb{N}$, are the intervals
\begin{eqnarray}
J_{n}:=[\,r_{n-1},r_{n}\,]\ \mathrm{with}\ r_{0}=0,\ r_{N}=R,\
\lambda_{n}:=r_{n}-r_{n-1}.
\end{eqnarray}
Following the finite element philosophy, we use bases of the piecewise polynomial function spaces that only comprise locally supported (global) basis functions. They are assembled from the following \emph{local shape functions} (see figure~\ref{fig:plot-01}):
\begin{eqnarray}
\hspace{-1.5cm}E^{0}_{n}(r)&:=& \left\{\,\begin{array}{r|l} \displaystyle 1\ &\ \mathrm{for}\ r\in J_{n}\\[5pt] \displaystyle 0\ &\ \mathrm{otherwise}
\end{array}\right.\\[5pt]
\hspace{-1.5cm}E^{1}_{n}(r)&:=& \left\{\,\begin{array}{r|l} \displaystyle\frac{1}{\lambda_{n}}\,\left(r-r_{n}\right)+1\ &\ \mathrm{for}\ r\in(r_{n-1},r_{n})\\[5pt] \displaystyle \frac{1+\delta_{nN}}{2}\ &\ \mathrm{for}\ r=r_{n}\\[5pt] \displaystyle 0\ &\ \mathrm{otherwise}\end{array}\right.\nonumber\\[5pt]
\hspace{-1.5cm}E^{2}_{n}(r)&:=& \left\{\,\begin{array}{r|l} \displaystyle\frac{1}{\lambda_{n}}\,\left(r_{n}-\lambda_{n}-r\right)+1\ &\ \mathrm{for}\ r\in(r_{n-1},r_{n})\\[5pt] \displaystyle \frac{1+\delta_{n0}}{2}\ &\ \mathrm{for}\ r=r_{n-1}\\[5pt] \displaystyle 0\ &\ \mathrm{otherwise}\end{array}\right.\nonumber\\[5pt]
\hspace{-1.5cm}E^{3}_{n}(r)&:=& \left\{\,\begin{array}{r|l} \displaystyle\frac{4}{\lambda^{2}_{n}}\,\left(r-r_{n}\right)\left(r_{n}-\lambda_{n}-r\right)\ &\ \mathrm{for}\ r\in J_{n}\\[5pt] \displaystyle 0\ &\ \mathrm{otherwise}\end{array}\right.\nonumber\\[5pt]
\hspace{-1.5cm}E^{4}_{n}(r)&:=& \left\{\,\begin{array}{r|l} \displaystyle\frac{64}{3\lambda^{3}_{n}}\,\left(r-r_{n}\right)\left(r_{n}-\lambda_{n}-r\right)\left(r_{n}-\frac{\lambda_{n}}{2}-r\right)\ &\ \mathrm{for}\ r\in J_{n}\\[5pt] \displaystyle 0\ &\ \mathrm{otherwise}
\end{array}\right.\nonumber
\end{eqnarray}
The parameters $N$, $r_{n}$ and $\lambda_{n}$ characterizing the mesh can be redefined at every time step (See below for the details of our adaption algorithm). The \emph{global shape functions} are obtained from the local shape functions through
\begin{eqnarray}
e^{B}_{i}(r):=\sum^{4,N}_{s,n=1}T^{Bn}_{is}E^{s}_{n}(r)
\end{eqnarray}
where $T^{Bn}_{is}$ are the combination coefficients
\begin{eqnarray}
T^{0n}_{is}&:=&\delta^{0}_{s}\delta^{n}_{i}\\[5pt]
T^{1n}_{is}&:=&\delta^{1}_{s}\delta^{n+1}_{i}+\delta^{2}_{s}\delta^{n}_{i}\nonumber\\[5pt]
T^{2n}_{is}&:=&\delta^{1}_{s}\delta^{2n+1}_{i}+\delta^{2}_{s}\delta^{2n-1}_{i}+\delta^{3}_{s}\delta^{2n}_{i}\nonumber\\[5pt]
T^{3n}_{is}&:=&\delta^{1}_{s}\delta^{3n+1}_{i}+\delta^{2}_{s}\delta^{3n-2}_{i}+\delta^{3}_{s}\delta^{3n-1}_{i}+\delta^{4}_{s}\delta^{3n}_{i}.\nonumber
\end{eqnarray}
With this definitions the sets $S_{B}:=\{e^{B}_{i}\}$ of global basis functions contain $N$ compactly supported functions of polynomial degree $p=0$ for $B=0$ and $BN+1$ compactly supported functions of polynomial degrees $p\in\{1,\ldots,B\}$ for $B\in\{1,\ldots,3\}$. For the finite element space with basis $S_{B}$ we adopt the notation $\mathcal{S}_{B}$. This spaces will play the role of the abstract {\sc Galerkin} trial/test space $V_{n}$ from sect.~\ref{sec:cons-disc}.
%
\subsection{The discrete equations}
\label{sec:discrete-equations}
\paragraph{} We represent the variables $X_{a}$, $Y_{a}$, $X_{b}$ and $Y_{b}$ of the {\sc Dirac} fields due to the discussion of section \ref{sec:cons-disc} as in (\ref{eq:Repralphabeta}) by functions from $\mathcal{S}_{B}$ for some fixed parameter $B\in\{1,\ldots,3\}$. For the variables $a$ and $b$ we just use a linear approximation by functions in $\mathcal{S}_{1}$:
\begin{eqnarray}
a(t_{m},r)\approx\sum_{i=1}^{N+1}a^{m,i}e^{1}_{i}\quad \mathrm{and}\quad b(t_{m},r)\approx\sum_{i=1}^{N+1}b^{m,i}e^{1}_{i}
\end{eqnarray}
The representations have to be compatible with the boundary conditions in (\ref{eq:InitValProb}), which implies
\begin{eqnarray}
X_{a}^{m,1}=Y_{a}^{m,1}=X_{b}^{m,1}=Y_{b}^{m,1}=0\\[5pt]
X_{a}^{m,(BN+1)}=Y_{a}^{m,(BN+1)}=X_{b}^{m,(BN+1)}=Y_{b}^{m,(BN+1)}=0\nonumber\\[5pt]
b^{m,1}=0\ \ \mathrm{and}\ \ a^{m,N+1}=-\,b^{m,N+1}.\nonumber
\end{eqnarray}
We construct the discretization of a weak formulation of the equations (\ref{eq:Xat}-\ref{eq:bt}) analogous to (\ref{eq:ModelProblemWeak}), whereby we have to use test functions from $\mathcal{S}_{0}$ for a stable discretization of the ODEs (\ref{eq:ar},\ref{eq:br}). We define the coefficients
\begin{eqnarray}
\begin{array}{rclrcl}
\fl A_{ij}&:=&\left(e^{B}_{i},e^{B}_{j}\right)\ &\ \displaystyle F_{ij}&:=&\left(\frac{1}{r}\,e^{1}_{i},e^{0}_{j}\right)\\[5pt]
\fl B_{ijk}&:=&\left(e^{1}_{i}\left(e^{B}_{j}\right)'+\frac{1}{2}\,\left(e^{1}_{i}\right)'e^{B}_{j},e^{B}_{k}\right)\ &\ \displaystyle G_{ijk}&:=&\left(\frac{1}{r}\left[\,e^{B}_{i}\left(e^{B}_{j}\right)'-\left(e^{B}_{i}\right)'e_{j}\,\right],e^{0}_{k}\right)\\[5pt]
\fl C_{ijk}&:=&\left(e^{1}_{i}e^{B}_{j},e^{B}_{k}\right)\ &\ \displaystyle H_{ijkl}&:=&\left(\frac{1}{r}\,e^{1}_{i}e^{B}_{j}e^{B}_{k},e^{0}_{l}\right)\\[5pt]
\fl D_{ijk}&:=&\left(\frac{1}{r}\,e^{1}_{i}e^{B}_{j},e^{B}_{k}\right)\ &\ \displaystyle I_{ijkl}&:=&\left(\frac{1}{r^{2}}\,e^{1}_{i}e^{B}_{j}e^{B}_{k},e^{0}_{l}\right)\\[5pt]
\fl E_{ij}&:=&\left(\left(e^{B}_{i}\right)',e^{0}_{j}\right)\ &\ \displaystyle J_{ijkl}&:=&\left(\frac{1}{r}\,e^{1}_{i}\left[\,\left(e^{B}_{j}\right)'e^{B}_{k}-e^{B}_{j}\left(e^{B}_{k}\right)'\,\right],e^{1}_{l}\right).
\end{array}
\end{eqnarray}
and finally arrive at the following discretized versions of (\ref{eq:Xat}-\ref{eq:bt})\footnote{Here we omitted the summation sign to shorten the notation. Analogous to {\sc Einstein}'s summation convention, we sum over the indices $i$, $j$ and $k$ where they appear twice.}:
\numparts\begin{eqnarray}
\fl 0&=&\frac{1}{\tau}\left[\,X_{a}^{m,i}-X_{a}^{m+1,i}\,\right]A_{is}+\frac{1}{2}\,\mathrm{e}^{\frac{\displaystyle a^{m,i}-b^{m,i}+a^{m+1,i}-b^{m+1,i}}{4}}\left[\,Y_{b}^{m,j}+Y_{b}^{m+1,j}\,\right]B_{ijs}\nonumber\\[7pt]
\fl {}&{}&{}\displaystyle+\,\frac{m}{2}\,\mathrm{e}^{\frac{\displaystyle a^{m,i}+a^{m+1,i}}{4}}\left[\,Y_{a}^{m,j}+Y_{a}^{m+1,j}\,\right]C_{ijs}+\,\frac{1}{2}\,\mathrm{e}^{\frac{\displaystyle a^{m,i}+a^{m+1,i}}{4}}\left[\,Y_{b}^{m,j}+Y_{b}^{m+1,j}\,\right]D_{ijs}\nonumber\\[7pt]
\fl 0&=&\frac{1}{\tau}\left[\,Y_{a}^{m,i}-Y_{a}^{m+1,i}\,\right]A_{is}-\frac{1}{2}\,\mathrm{e}^{\frac{\displaystyle a^{m,i}-b^{m,i}+a^{m+1,i}-b^{m+1,i}}{4}}\left[\,X_{b}^{m,j}+X_{b}^{m+1,j}\,\right]B_{ijs}\nonumber\\[7pt]
\fl {}&{}&{}\displaystyle-\,\frac{m}{2}\,\mathrm{e}^{\frac{\displaystyle a^{m,i}+a^{m+1,i}}{4}}\left[\,X_{a}^{m,j}+X_{a}^{m+1,j}\,\right]C_{ijs}-\,\frac{1}{2}\,\mathrm{e}^{\frac{\displaystyle a^{m,i}+a^{m+1,i}}{4}}\left[\,X_{b}^{m,j}+X_{b}^{m+1,j}\,\right]D_{ijs}\nonumber\\[7pt]
\label{eq:DiracDisc}\fl 0&=&\frac{1}{\tau}\left[\,X_{b}^{m,i}-X_{b}^{m+1,i}\,\right]A_{is}-\frac{1}{2}\,\mathrm{e}^{\frac{\displaystyle a^{m,i}-b^{m,i}+a^{m+1,i}-b^{m+1,i}}{4}}\left[\,Y_{a}^{m,j}+Y_{a}^{m+1,j}\,\right]B_{ijs}\\[7pt]
\fl {}&{}&{}\displaystyle-\,\frac{m}{2}\,\mathrm{e}^{\frac{\displaystyle a^{m,i}+a^{m+1,i}}{4}}\left[\,Y_{b}^{m,j}+Y_{b}^{m+1,j}\,\right]C_{ijs}+\,\frac{1}{2}\,\mathrm{e}^{\frac{\displaystyle a^{m,i}+a^{m+1,i}}{4}}\left[\,Y_{a}^{m,j}+Y_{a}^{m+1,j}\,\right]D_{ijs}\nonumber\\[7pt]
\fl 0&=&\frac{1}{\tau}\left[\,Y_{b}^{m,i}-Y_{b}^{m+1,i}\,\right]A_{is}+\frac{1}{2}\,\mathrm{e}^{\frac{\displaystyle a^{m,i}-b^{m,i}+a^{m+1,i}-b^{m+1,i}}{4}}\left[\,X_{a}^{m,j}+X_{a}^{m+1,j}\,\right]B_{ijs}\nonumber\\[7pt]
\fl {}&{}&{}\displaystyle+\,\frac{m}{2}\,\mathrm{e}^{\frac{\displaystyle a^{m,i}+a^{m+1,i}}{4}}\left[\,X_{b}^{m,j}+X_{b}^{m+1,j}\,\right]C_{ijs}-\,\frac{1}{2}\,\mathrm{e}^{\frac{\displaystyle a^{m,i}+a^{m+1,i}}{4}}\left[\,X_{a}^{m,j}+X_{a}^{m+1,j}\,\right]D_{ijs}\nonumber
\end{eqnarray}
\begin{eqnarray}
\label{eq:arDisc}\fl 0&=&-\,a^{m+1,i}E_{is}+\left[\,\mathrm{e}^{b^{m+1,i}}-1\,\right]F_{is}+4\,\left[\,X_{a}^{m+1,i}X_{b}^{m+1,j}+Y_{a}^{m+1,i}Y_{b}^{m+1,j}\,\right]G_{ijs}
\end{eqnarray}
\begin{eqnarray}
\label{eq:brDisc}\fl 0&=&-\,b^{m+1,i}E_{is}+\left[\,1-\mathrm{e}^{b^{m+1,i}}\,\right]F_{is}+4\,\left[\,X_{a}^{m+1,i}X_{b}^{m+1,j}+Y_{a}^{m+1,i}Y_{b}^{m+1,j}\,\right]G_{ijs}\\[7pt]
\fl
{}&{}&\displaystyle+\,4m\,\mathrm{e}^{\frac{b^{m+1,i}}{2}}\,\left[\,X_{a}^{m+1,j}X_{a}^{m+1,k}+Y_{a}^{m+1,j}Y_{a}^{m+1,k}-X_{b}^{m+1,j}X_{b}^{m+1,k}-Y_{b}^{m+1,j}Y_{b}^{m+1,k}\,\right]H_{ijks}\nonumber\\[7pt]
\fl
{}&{}&\displaystyle+\,8\,\mathrm{e}^{\frac{b^{m+1,i}}{2}}\,\left[\,X_{a}^{m+1,j}X_{b}^{m+1,k}+Y_{a}^{m+1,j}Y_{b}^{m+1,k}\,\right]I_{ijks}\nonumber
\end{eqnarray}
\begin{eqnarray}
\label{eq:btDisc}\fl 0&=& \frac{1}{\tau}\left[\,b^{m,i}-b^{m+1,i}\,\right]A_{is}+\mathrm{e}^{\frac{\displaystyle a^{m,i}-b^{m,i}+a^{m+1,i}-b^{m+1,i}}{4}}\\[7pt]
\fl
{}&{}&\displaystyle\cdot\left[\,\left[\,X_{a}^{m,j}+X_{a}^{m+1,j}\,\right]\left[\,Y_{a}^{m,k}+Y_{a}^{m+1,k}\,\right]+\left[\,X_{b}^{m,j}+X_{b}^{m+1,j}\,\right]\left[\,Y_{b}^{m,k}+Y_{b}^{m+1,k}\,\right]\,\right]J_{ijks}\nonumber
\end{eqnarray}\endnumparts
Note that although we are free to decide which of the equations (\ref{eq:Br}) or (\ref{eq:Bt}) we consider as redundant, the discrete systems (\ref{eq:DiracDisc},\ref{eq:arDisc},\ref{eq:brDisc}) and (\ref{eq:DiracDisc},\ref{eq:arDisc},\ref{eq:btDisc}) are not exactly equivalent!
%
\subsection{The algorithms}
\label{sec:algorithms}
\paragraph{} The following algorithms have been implemented in MATLAB:
\begin{enumerate}
\item An initial mesh can be given, initial conditions can be chosen from the families (\ref{eq:InitCondVen}) or (\ref{eq:InitCondStd}) and the software projects them in $L^{2}(J_{R})$ to a linear combination of the form (\ref{eq:Repralphabeta}). All necessary integrals are evaluated by {\sc Gaussian} quadrature \cite{Haemmerlin_Hoffmann_NumMath}, whose order can be specified in the parameter $P_{\mathrm{Go}}$.
\item The initial metric is obtained by solving the equations (\ref{eq:arDisc},\ref{eq:brDisc}) implicitly by applying {\sc Newton}'s algorithm \cite{Haemmerlin_Hoffmann_NumMath} until the equations are satisfied up to a specified residuum $P_{\mathrm{Ntol}}$ or a given maximum number $P_{\mathrm{Nmax}}$ of iterations is reached.
\item Before each time step the mesh is updated due to the following procedure: The initial mesh structure is represented by a density function $\rho(r)$, such that for all $u,v\in J_{R}$
\begin{eqnarray}
\int_{u}^{v}\rho\,dr=\mathrm{Number\ of\ cells\ in}\ [\,u,v\,].
\end{eqnarray}
The automatic mesh update ensures
\begin{eqnarray}
\int_{u}^{v}\rho\,\sqrt{B}\,dr=\mathrm{Number\ of\ cells\ in}\ [\,u,v\,]
\end{eqnarray}
by dividing a cell into two cells if
\begin{eqnarray}
\int_{r_{j-1}}^{r_{j}}\rho\,\sqrt{B}\,dr\approx\frac{\lambda_{j}}{2}\,\left[\rho^{j-1}\mathrm{e}^{\frac{b^{j-1}}{2}}+\rho^{j}\mathrm{e}^{\frac{b^{j}}{2}}\right]>P_{\mathrm{ath}},
\end{eqnarray}
where $P_{\mathrm{ath}}$ is a threshold that can be specified as a parameter. The above criterion means the number of cells per physical arclength in radial direction is given by the initially specified function $\rho$ on each spacelike slice $\Sigma(t)$. This was motivated by the principle of equivalence.
\item If the size $\tau$ of the time steps is chosen to be too large, the discrete domain of dependence does not encompass that of the continuum system and the resulting time evolution may yield spurious solutions. Therefore we choose
\begin{eqnarray}
\tau:=t_{m+1}-t_{m}=P_{\lambda\tau}\cdot\min_{j}\left(\lambda_{j}\mathrm{e}^{\frac{b^{m,j}-a^{m,j}}{2}}\right),
\end{eqnarray}
where $P_{\lambda\tau}<1$ can be specified as a parameter.
\item For each time step the equations (\ref{eq:DiracDisc}-\ref{eq:brDisc}) are solved implicitly by applying {\sc Newton}'s algorithm until the residuum drops below a specified threshold $P_{\mathrm{Ntol}}$ or the given maximum number $P_{\mathrm{Nmax}}$ of iterations is reached. In the latter case the time step is rejected if the residuum exceeds $P_{\mathrm{Nmaxtol}}$ and repeated with a smaller $\tau_{\mathrm{new}}:=P_{\mathrm{a\tau\lambda}}\cdot\tau_{\mathrm{old}}$.  After $P_{\mathrm{maxtry}}$ refinements of $\tau$ the program exits with an error message.
\item The residues of the equations (\ref{eq:DiracDisc}-\ref{eq:brDisc}) and (\ref{eq:btDisc}) as well as the change in the total electric charge and the ADM-mass are monitored and thus, can be used for accuracy checks.
\end{enumerate}
%
\section{Numerical experiments}
\label{sec:numer-exper}
%
\subsection{Convergence Tests}
\label{sec:conver-test}
\paragraph{} In order to study the convergence of the scheme, we conducted simulations on an equidistant static meshgrid, i.e. with automatic meshgrid and timestep refinement turned off. We performed $9$ runs due to all possible combinations of the parameters
\begin{eqnarray}
B\in\{1,\ldots,3\}\ \mathrm{ and }\ N\in\{120,240,480\}
\end{eqnarray}
for the initial data family (\ref{eq:InitCondVen}) using the fixed parameter values
\begin{eqnarray}
P_{\mathrm{Go}}=150,\ P_{\mathrm{Ntol}}=9\cdot 10^{-\,16},\ P_{\mathrm{Nmaxtol}}=10^{-\,13},\ P_{\mathrm{Nmax}}=30\nonumber\\[5pt] 
P_{\mathrm{ath}}=1.25,\ P_{\lambda\tau}=0.1,\ P_{\mathrm{a\tau\lambda}}=0.5,\ P_{\mathrm{maxtry}}=50,\\[5pt]
m=0.25,\ \Sigma=0.3,\ R=5\ \mathrm{ and }\ T=3.125.\nonumber
\end{eqnarray}
Let $u_{(B,N)}(r)$ be the numerical solution from the $(B,N)$-run for one of the variables in $\{X_{a},Y_{a},X_{b},Y_{b},a,b\}$ at fixed time $T$ and $u(r)$ denotes the corresponding exact solution. Due to {\sc Richardson}'s extrapolation, we assume there is a function $f_{B}(r)$ and a $p_{B}>0$, such that
\begin{eqnarray}
u=u_{(B,N)}+f_{B}\cdot (\lambda_{N})^{p_{B}},
\end{eqnarray}
where $\lambda_{N}:=\frac{R}{N}$ is the grid spacing. For each run, we evaluated $u_{(B,N)}$ on an equidistant test-grid with $N_{T}:=10000$ gridpoints, which yields the values $\left\{u_{(B,N)}^{(n)}\right\}$ for $n\in\{1,\ldots,N_{T}\}$ and computed the approximation
\begin{eqnarray}
p_{B}\approx\mathrm{mean}_{n\in\{1,\ldots,N_{T}\}}\left(\left.\ln\left(\,\left\vert\frac{u_{(B,120)}^{(n)}-u_{(B,240)}^{(n)}}{u_{(B,240)}^{(n)}-u_{(B,480)}^{(n)}}\right\vert\,\right)\right\slash\ln(2)\right).
\end{eqnarray}
The resulting values for $p_{B}$ are summarized in the following table:\\
\begin{center}
\begin{tabular}{cccccccccc}\hline
B&$X_{a}$&$Y_{a}$&$X_{b}$&$Y_{b}$&$a$&$b$&$\mathrm{mean}_{\left\{X_{a},\ldots, Y_{b}\right\}}$&$\mathrm{mean}_{\left\{a,b\right\}}$&mean all\\\hline
$1$&$2.73$&$2.25$&$4.77$&$2.59$&$1.94$&$1.95$&$3.08$&$1.95$&$2.70$\\
$2$&$1.95$&$2.13$&$1.96$&$1.90$&$3.13$&$3.05$&$1.98$&$3.09$&$2.35$\\
$3$&$3.06$&$2.91$&$5.48$&$2.83$&$2.15$&$2.17$&$3.57$&$2.16$&$3.10$\\\hline
\end{tabular}
\end{center}
%
\subsection{Experiments near the blackhole treshold}
\label{sec:exp-near-bht}
\begin{figure}[tp]
\begin{center}
\begin{picture}(15,21)
\put(0,0){\includegraphics[width=16cm,height=21cm]{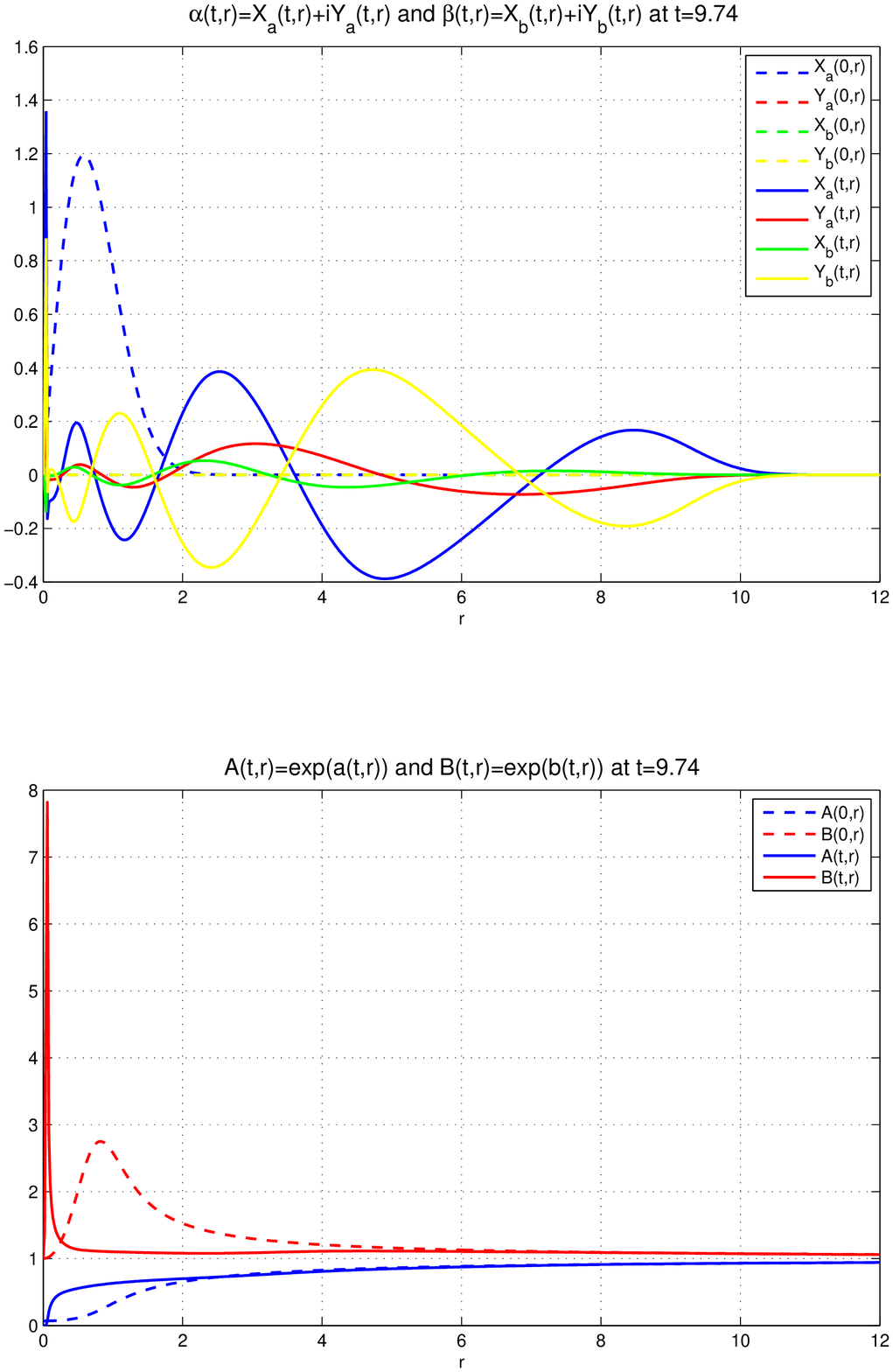}}
\end{picture}
{\bf \caption{\label{fig:plot-02}}\footnotesize {\textmd {Plot of the variables as functions of $r$ for the run with $R=12$, $B=3$, $N_{0}=240$ and $\Sigma=0.41185$, after performing
$M=5150$ time steps.}}}
\end{center}
\end{figure}

\begin{figure}[tp]
\begin{center}
\begin{picture}(15,21)
\put(0,0){\includegraphics[width=16cm,height=21cm]{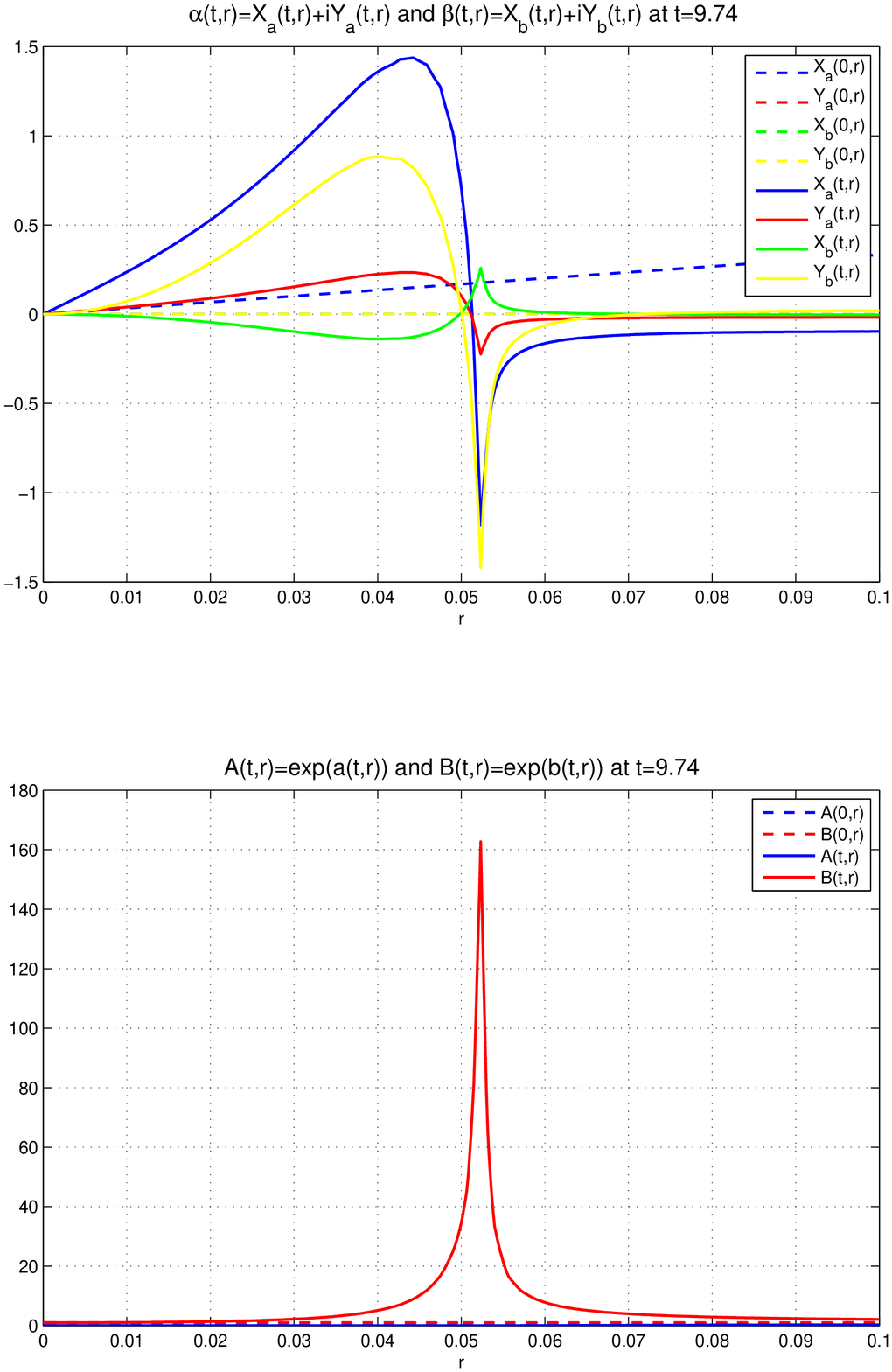}}
\end{picture}
{\bf \caption{\label{fig:plot-03}}\footnotesize {\textmd {Figure \ref{fig:plot-02} zoomed to show the forming of the black hole.}}}
\end{center}
\end{figure}

\begin{figure}[tp]
\begin{center}
\begin{picture}(15,21)
\put(0,0){\includegraphics[width=16cm,height=21cm]{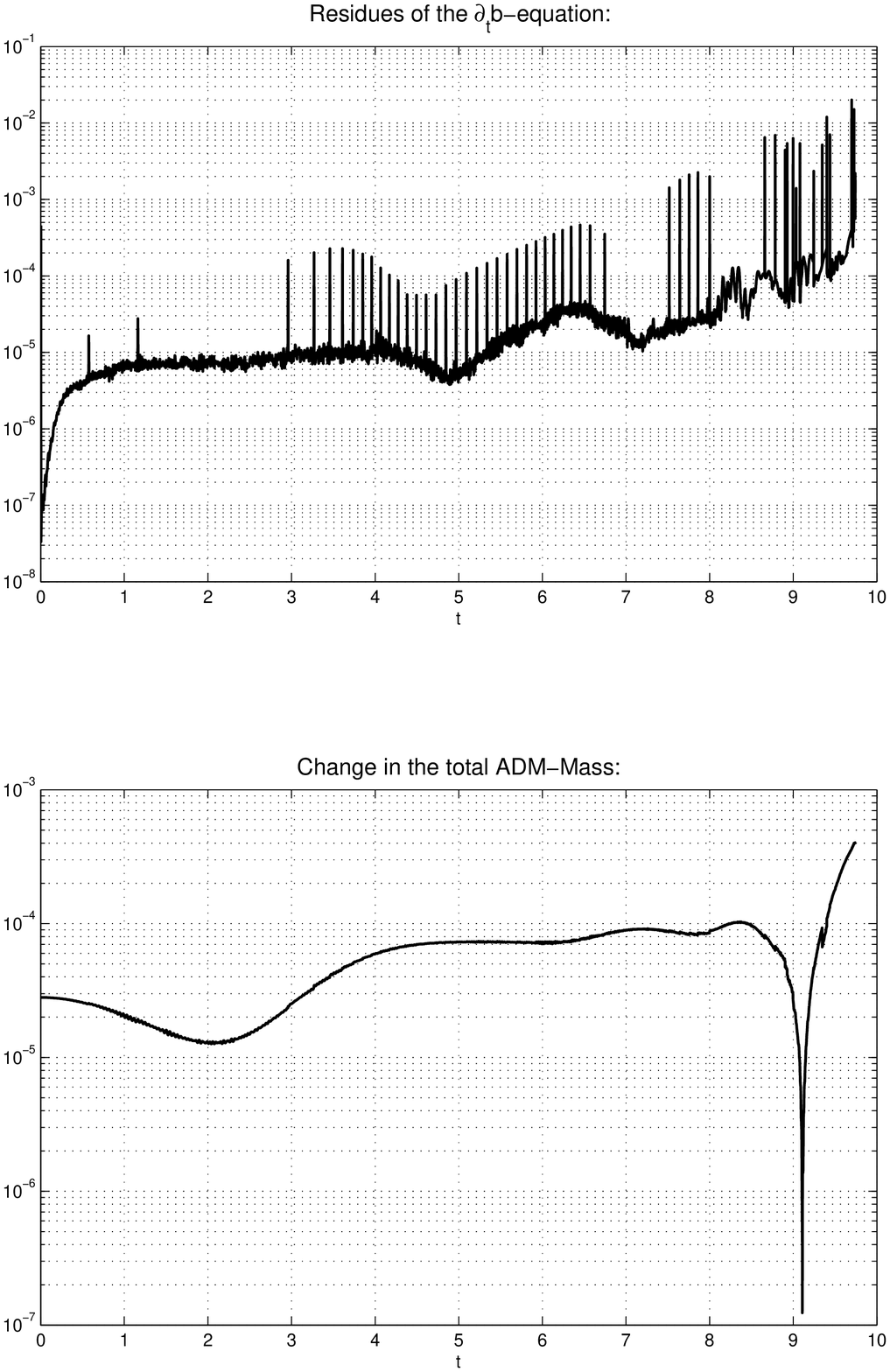}}
\end{picture}
{\bf \caption{\label{fig:plot-04}}\footnotesize {\textmd{Residues of the redundant equation (\ref{eq:btDisc}) and drift of the ADM-Mass for the run with $R=12$, $B=3$, $N_{0}=240$ and $\Sigma=0.41185$.}}}
\end{center}
\end{figure}

\begin{figure}[tp]
\begin{center}
\begin{picture}(15,21)
\put(0,0){\includegraphics[width=16cm,height=21cm]{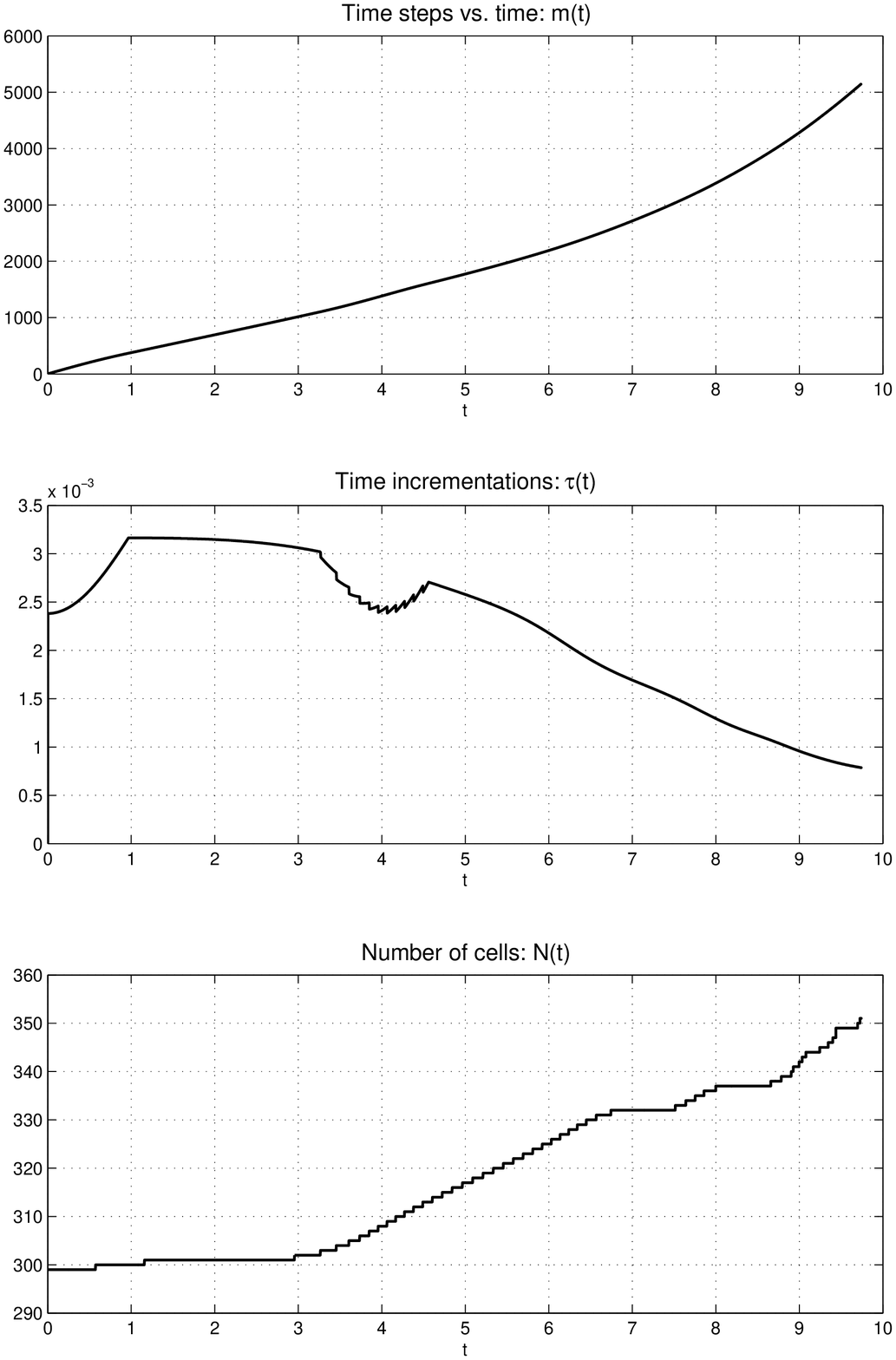}}
\end{picture}
{\bf \caption{\label{fig:plot-05}}\footnotesize {\textmd{Development of the mesh due to our adaption algorithms for the run with $R=12$, $B=3$, $N_{0}=240$ and $\Sigma=0.41185$.}}}
\end{center}
\end{figure}

\begin{figure}[tp]
\begin{center}
\begin{picture}(15,21)
\put(0,0){\includegraphics[width=16cm,height=21cm]{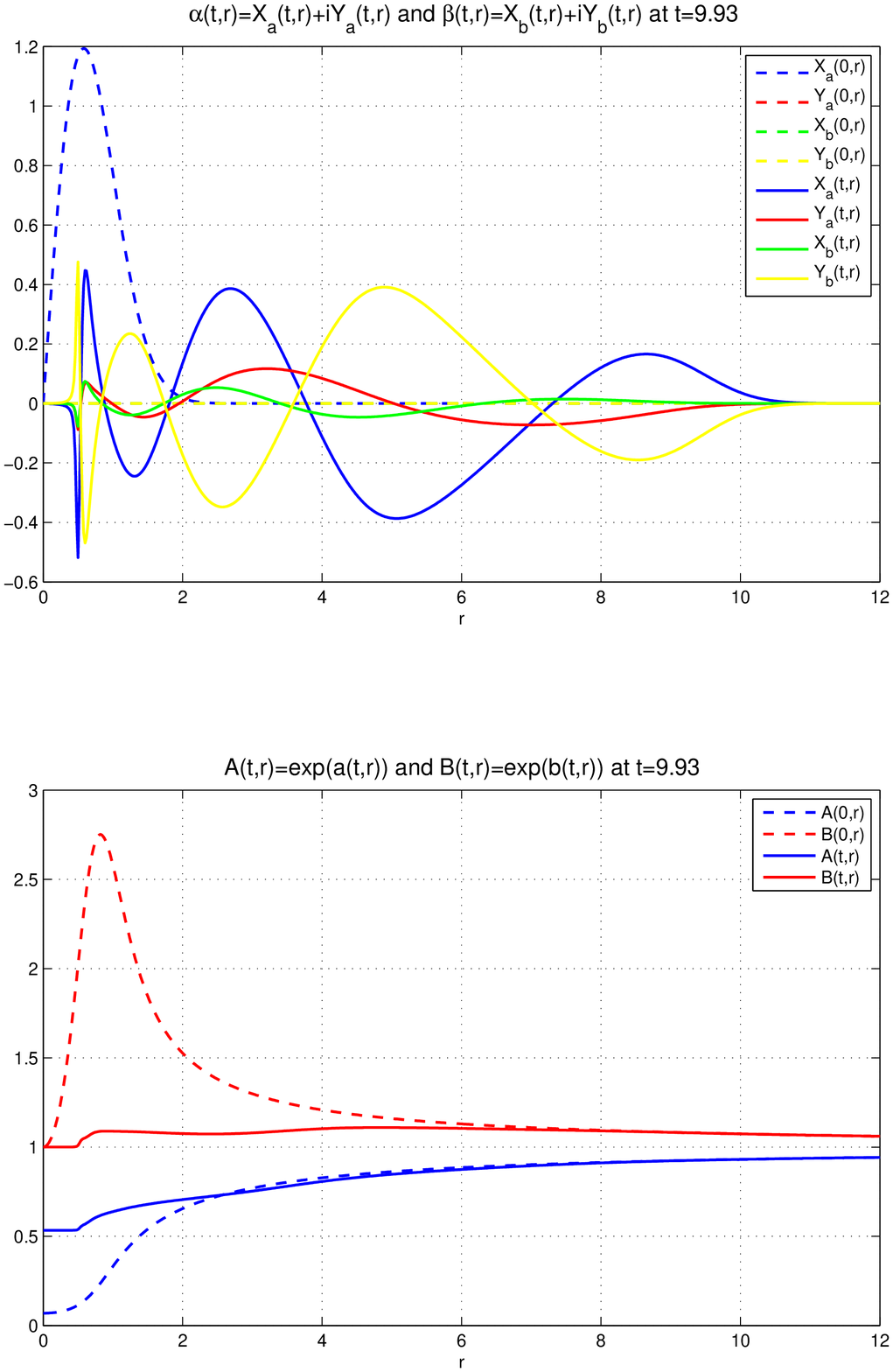}}
\end{picture}
{\bf \caption{\label{fig:plot-06}}\footnotesize {\textmd {Plot of the variables as functions of $r$ for the run with $R=12$, $B=3$, $N_{0}=240$ and $\Sigma=0.41186$, after performing $M=5950$ time steps.}}}
\end{center}
\end{figure}

\begin{figure}[tp]
\begin{center}
\begin{picture}(15,21)
\put(0,0){\includegraphics[width=16cm,height=21cm]{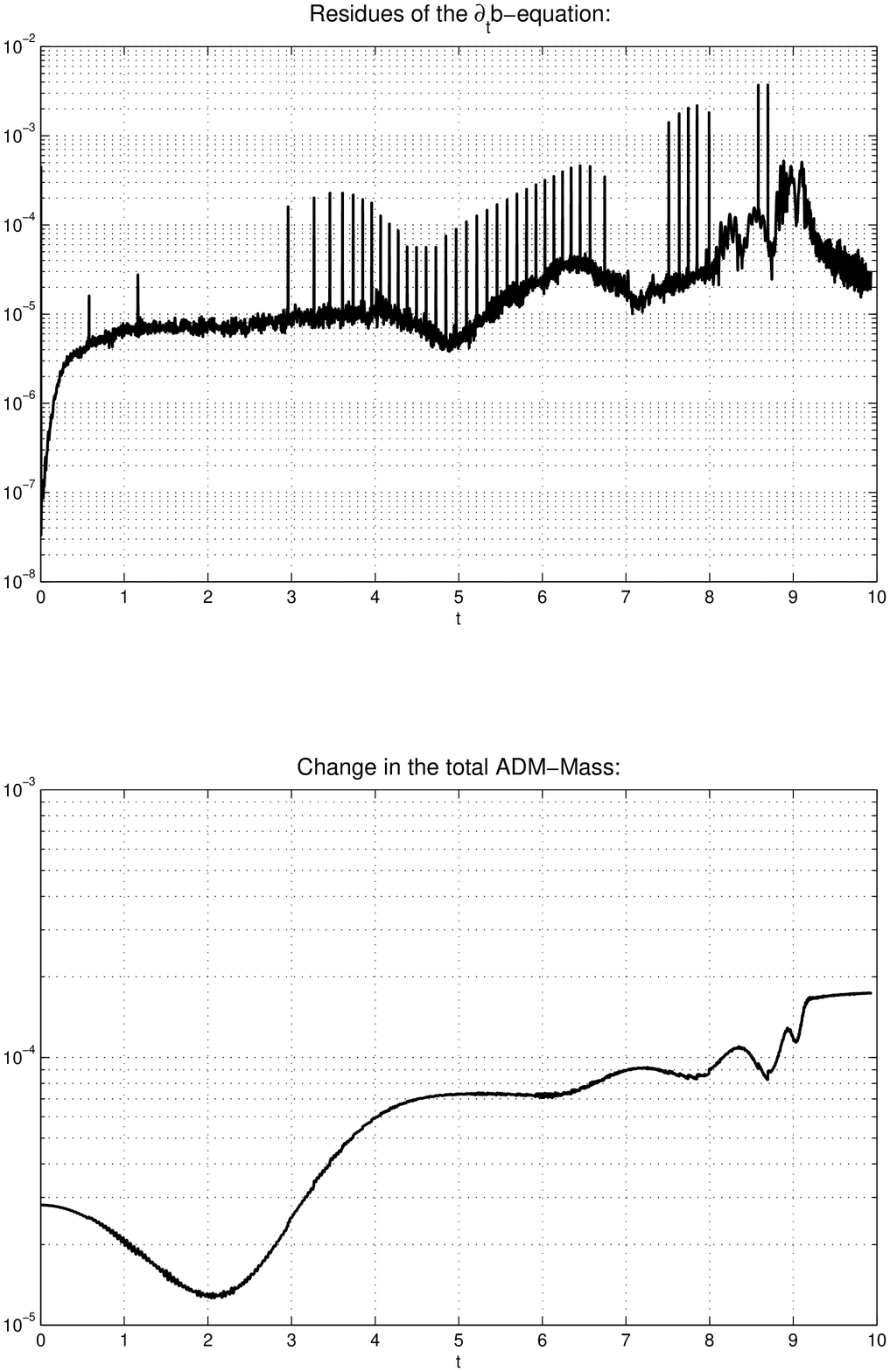}}
\end{picture}
{\bf \caption{\label{fig:plot-07}}\footnotesize {\textmd {Residues of the redundant equation (\ref{eq:btDisc}) and drift of the ADM-Mass for the run with $R=12$, $B=3$, $N_{0}=240$ and $\Sigma=0.41186$.}}}
\end{center}
\end{figure}

\begin{figure}[tp]
\begin{center}
\begin{picture}(15,21)
\put(0,0){\includegraphics[width=16cm,height=21cm]{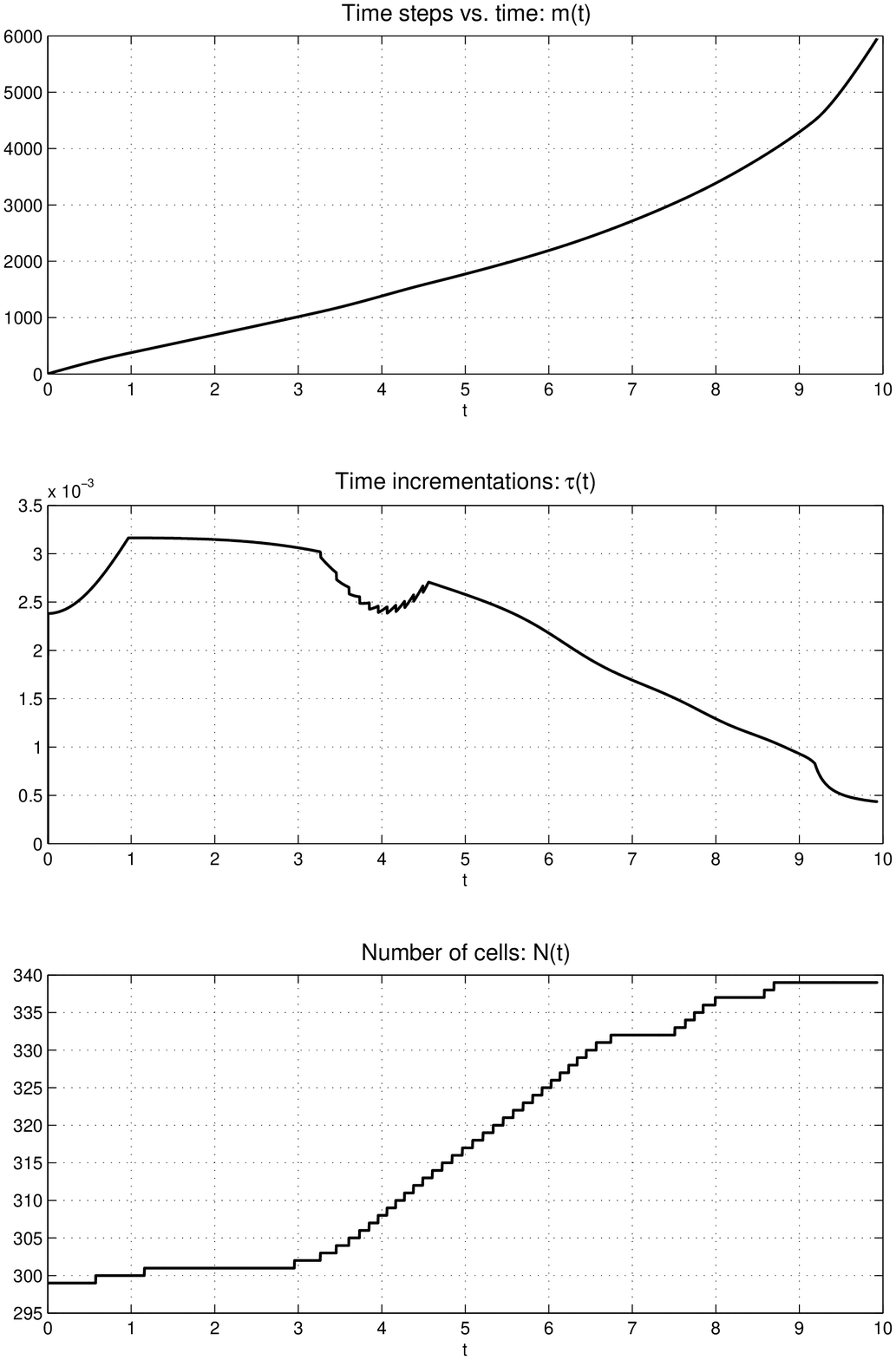}}
\end{picture}
{\bf \caption{\label{fig:plot-08}}\footnotesize {\textmd {Development of the mesh due to our adaption algorithms for the run with $R=12$, $B=3$, $N_{0}=240$ and $\Sigma=0.41186$.}}}
\end{center}
\end{figure}

\paragraph{} We tested our code for the initial data family (\ref{eq:InitCondVen}) at particle mass $m=0.25$ and compared our results to \cite{Schaefer_Steck_Ventrella_CritCollotEDF}. We specified the initial mesh by the cell distribution function
\begin{eqnarray}
D_{\mathrm{c}}^{k}:=1+P_{\mathrm{Dc}}\left[\,1-\cos\left(\frac{k\pi}{N_{0}}\right)\,\right],
\end{eqnarray}
from which the cell widths and the density function are computed by
\begin{eqnarray}
\lambda_{k}:=R\,\frac{D_{\mathrm{c}}^{k}}{\sum_{j=1}^{N_{0}}D_{\mathrm{c}}^{j}}\ \mathrm{and}\ \rho^{k}:=\frac{1}{\lambda_{k}}\,.
\end{eqnarray}
This results in an equidistant grid for $P_{\mathrm{Dc}}=0$ and for $P_{\mathrm{Dc}}>0$ the density of cells decreases with increasing $r$. For all runs we specified the following fixed parameter values:
\begin{eqnarray}
P_{\mathrm{Go}}=150,\ P_{\mathrm{Ntol}}=9\cdot 10^{-\,16},\ P_{\mathrm{Nmaxtol}}=10^{-\,13},\ P_{\mathrm{Nmax}}=30\\[5pt] 
P_{\mathrm{ath}}=1.25,\ P_{\lambda\tau}=0.1,\ P_{\mathrm{a\tau\lambda}}=0.5,\ P_{\mathrm{maxtry}}=50,\ P_{\mathrm{Dc}}=7.\nonumber
\end{eqnarray}
The remaining parameters to be specified for each run are $m$, $R$, $B$, $N_{0}$ and $\Sigma$. For $\Sigma$ small enough, the system collapsed to form a black hole, whereas all matter travels toward spatial infinity for large $\Sigma$. Using {\sc Crank-Nicholson} finite differencing, {\sc Schaefer, Steck} and {\sc Ventrella} found the physical threshold $\Sigma_{\mathrm{th}}$ near $\Sigma_{\mathrm{thSSV}}:=0.412$ for $m=0.25$ in \cite{Schaefer_Steck_Ventrella_CritCollotEDF}. 
\paragraph{} The decision, whether a black hole emerged, was based on the existence of a $t$ and $r$, such that
\begin{eqnarray}
F(t,r):=\frac{2M(t,r)}{r}=1-\frac{1}{B(t,r)}>1-\epsilon_{\mathrm{BH}}
\end{eqnarray}
for a small parameter $\epsilon_{\mathrm{BH}}>0$. By carefully varying $\Sigma$, we found the numerical blackhole threshold
$\Sigma_{\mathrm{thZ}}$ to satisfy
\begin{eqnarray}
0.41185<\Sigma_{\mathrm{thZ}}<0.41186
\end{eqnarray}
for runs using $B=3$ and $N_{0}=240$. This confirms\footnote{Since we don't know anything about the error with respect to the exact value of the treshold, we called $\Sigma_{\mathrm{thZ}}$ the ``numerical" treshold and publish all of its digits found. As mentioned above, {\sc Schaefer, Steck} and {\sc Ventrella} reported the critical value of $\Sigma_{\mathrm{thSSV}}=0.412$, which was rounded to $3$ significant digits, but was actually found to higher accuracy. Our result confirms at least this $3$ digits published in \cite{Schaefer_Steck_Ventrella_CritCollotEDF}.} the result of
\cite{Schaefer_Steck_Ventrella_CritCollotEDF}. We show the six variables as a function of $r$ at a fixed value of $t$ at the end of the run with $\Sigma=0.41185$ in figure \ref{fig:plot-02}. To investigate the details at the forming black hole horizon, we show a zoomed version in figure \ref{fig:plot-03}. In this run the black hole was detected at
\begin{eqnarray}
R_{\mathrm{BH}}\approx 0.052,\ M_{\mathrm{BH}}:=\frac{R_{\mathrm{BH}}}{2}\approx 0.026\ \ \mathrm{using}\ \ \epsilon_{\mathrm{BH}}:=0.9937.
\end{eqnarray}
The residue (difference of the right hand side from $0$) of the redundant equation (\ref{eq:btDisc}) was saved for each global shape function at each timestep. The maximum over the global shape functions at each time is plotted in figure \ref{fig:plot-04}, as well as the relative drift of the conserved total ADM-Mass, i.e.
\begin{eqnarray}
\mathrm{Err}_{M}(t):=\left\vert\,\frac{M_{\mathrm{ADM}}(t)-M_{\mathrm{ADM}}(0)}{M_{\mathrm{ADM}}(0)}\,\right\vert.
\end{eqnarray}
Finally we show the development of the mesh\footnote{Although we specified a value of $N_{0}=240$, the number of cells in the plot starts at $299$ because of the mesh adaption due to the initial metric, which takes place before the time development starts.} due to our adaption algorithms in figure \ref{fig:plot-05}. This consists of the number of timesteps performed to reach up to time $t$, the timestep width $\tau(t)$ and the Number of cells $N(t)$ as functions of $t$. By comparing these plots to their analogues for the run with $\Sigma=0.41186$ in figures \ref{fig:plot-06}-\ref{fig:plot-08}, we conclude, that runs near the black hole threshold are well behaved and can be well studied before the numerics break down or are slowed down due to exorbitant mesh adaption. Therefore our algorithm seems to be suitable for studying critical collapse.
%
%

%
\end{document}